\documentclass[12pt, aps, prx, onecolumn, superscriptaddress, notitlepage]{revtex4-1} 

\usepackage{amsmath,amssymb,amsfonts,bm}
\usepackage{mathtools}
\usepackage{graphicx}
\usepackage{xcolor}
\usepackage{bbold}
\usepackage{graphicx}
\usepackage{dcolumn}
\usepackage{epstopdf}
\usepackage{epsfig}
\usepackage[caption=false]{subfig}
\usepackage{url}
\usepackage{hyperref}
\usepackage{verbatim}
\usepackage[export]{adjustbox}
\usepackage{scalerel}
\usepackage{braket}

\usepackage{tikz}

\usepackage{amsthm}
\usepackage{enumitem}

\usepackage{multirow}
\usepackage{booktabs}
\usepackage{pifont}

\usepackage{colortbl}

\newcommand{\id}{\mathbb{1}}

\usepackage{float}
\newcommand{\Wg}{\operatorname{Wg}}

\newcommand{\uu}{u}
\newcommand{\UU}{U}

\newcommand{\haarU}{\mathcal{U}}

\newcommand{\be}{\begin{equation}}
\newcommand{\ee}{\end{equation}}
\newcommand{\ba}{\begin{aligned}}
\newcommand{\ea}{\end{aligned}}

\newcommand{\bmult}{\begin{multline}}
\newcommand{\emult}{\end{multline}}

\newcommand{\dist}[1]{\left| \left| {#1} \right| \right|}

\newcommand{\rhoH}{\rho_{\rm H}}
\newcommand{\eH}{\mathcal{E}_{\rm H}}
\newcommand{\FH}{F_{\rm H}}

\newcommand{\HH}{N}

\def\Tr{\operatorname{Tr}}

\def\EPE{\mathcal{E}_{\rm PE}}

\graphicspath{{figures/}}

\newcommand{\aaa}{\textbf{a}}

\newcommand{\kket}[1]{\ket{\!\ket{#1}\hspace{-2.7pt}}}
\newcommand{\bbra}[1]{\bra{\!\bra{#1}\hspace{-1.2pt}}}
\newcommand{\doublev}{|\hspace{-0.05cm}|}
\newcommand{\bbraket}[1]{\braket{\!\braket{#1}\hspace{-2.0pt}}}

\begin{document}

\title{Projected state ensemble of a generic model of many-body quantum chaos}

\author{Amos Chan}
\address{Department of Physics, Lancaster University, Lancaster LA1 4YB, United Kingdom}
\email{amos.chan@lancaster.ac.uk}

\author{Andrea De Luca}
\address{Laboratoire de Physique Th\'eorique et Mod\'elisation, CNRS UMR 8089,
	CY Cergy Paris Universit\'e, 95302 Cergy-Pontoise Cedex, France}
\email{andrea.de-luca@cyu.fr}

\begin{abstract}
The projected ensemble is based on the study of the quantum state of a subsystem $A$ conditioned on projective measurements in its complement. Recent studies have observed that a more refined measure of the thermalization of a chaotic quantum system can be defined on the basis of convergence of the projected ensemble to a quantum state design, i.e. a system thermalizes when it becomes indistinguishable, up to the $k$-th moment, from a Haar ensemble of uniformly distributed pure states. Here we consider a random unitary circuit with the brick-wall geometry and analyze its convergence to the Haar ensemble through the frame potential and its mapping to a statistical mechanical problem. This approach allows us to highlight a geometric interpretation of the frame potential based on the existence of a fluctuating membrane, similar to those appearing in the study of entanglement entropies. At large local Hilbert space dimension $q$, we find that all moments converge simultaneously with a time scaling linearly in the size of region $A$, a feature previously observed in dual unitary models. However, based on the geometric interpretation, we argue that the scaling at finite $q$ on the basis of rare membrane fluctuations, finding the logarithmic scaling of design times $t_k = O(\log k)$. Our results are supported with numerical simulations performed at $q=2$. 
\end{abstract}

%
%
%
%
%

\maketitle

\section{Introduction}
The problem of thermalization has been a central topic in the study of many-body systems in physics in recent years. In the framework of closed dynamics, the conventional understanding of thermalization is based on the idea that a large many-body system can act as its own thermal bath, leading to relaxation towards thermal equilibrium when any finite subportion is kept in contact with an infinitely large complementary system~\cite{cazalilla2010focus,polkovnikov2011colloquium,dalessio2016quantum}. This paradigm has been extensively studied and its origin has been related to the fundamental properties of the statistics of eigenstates, showing the importance of a probabilistic approach in complex quantum systems, ultimately leading to the use of random matrices~\cite{Mehta} and the notion of \emph{eigenstate thermalisation hypothesis}~\cite{deutsch1991,srednicki1994chaos,rigol2008thermalization}. The emergence of thermalisation has been verified in several numerical~\cite{PhysRevE.90.052105} and even experimental platforms~\cite{trotzky2012probing,langen2013local,geiger2014local,langen2015experimental,neill2016ergodic,clos2016time,kaufman2016quantum}. In most cases, the experimental observation of a subsystem is naturally associated to the \emph{measurement} of its complement, an information that is usually discarded in the standard approach to thermalisation. Thus, for both fundamental and practical motivations, it becomes interesting to explore 
a thermalization process when the bath is subject to projective measurements and the measurement outcomes are known and stored~\cite{PhysRevA.103.062218}. This scenario leads to the notion of \emph{projected ensemble} (PE) \cite{cotler2021emergentquantum,choi2021emergentrandomness}, where after preparing a state through unitary evolution, one focuses on the ensemble of pure states in the subsystem $A$ conditioned on the outcomes of quantum measurements in the complement $B$ (bath). The PE can be interpreted as a particular unraveling of the reduced density matrix, which is recovered if the information about $B$ is discarded. Thus, going beyond the characterisation of the reduced density matrix, one introduces \emph{deep thermalization}~\cite{ippoliti2022dynamical, shrotriya2023nonlocality}, whose goal, in broad terms, is to determine the minimal amount of information required to fully characterize this projected ensemble.
Arguing on the basis of ergodicity and typicality of many-body quantum systems, it is reasonable to assume that, for generic chaotic systems and in the absence of specific local conserved quantities,  the PE approaches a uniform distribution over the set of pure states in $A$, known as the \emph{Haar ensemble}. Such a natural conjecture has been confirmed by numerical and experimental evidence in~\cite{cotler2021emergentquantum,choi2021emergentrandomness}. More analytical and even rigorous results provided in Refs.~\cite{ho2022exact,claeys2022emergent,ippoliti2022dynamical} 
have further substantiated these claims for classes of strongly chaotic yet tractable models in $1D$, known as \emph{dual-unitary} circuits (DUC)~\cite{Bertini2018sff, bertini2019entanglement,bertini2019exact,piroli2020exact}.
The fact that an ensemble of pure states provides a good approximation of the uniform Haar distribution is an important characteristic known as \emph{quantum state design}~\cite{renes2004symmetric,ambainis2007quantum}. More specifically, consider an ensemble of pure states $\mathcal{E} = \{ p(i), \ket{\Psi(i)}\in \mathcal{H}\}_i$,
where $p(i)$ is the probability of picking the pure state $\ket{\Psi(i)}$, labeled by $i$, in the Hilbert space $\mathcal{H}$. Then, $\mathcal{E}$ is dubbed a $k$-design if it provides a good approximation, up to a small $\epsilon$, of the Haar distribution up to $k$ tensor powers of the state itself, i.e.
\begin{equation}
\label{eq:deepth}
    \sum_i p(i) (\ket{\Psi(i)}\bra{\Psi(i)})^{\otimes k} \sim \int_{\rm Haar} d\psi (\ket{\psi}\bra{\psi})^{\otimes k}  \;.
\end{equation}
Having efficient methods and protocols to approximate Haar-distributed states is useful to achieve a practical means of scalable randomized benchmarking to assess errors in quantum computing operations~\cite{Emerson_2005}. 
Fundamentally, the interest in the projected ensemble protocol lies in the inherent randomness of the outcomes of quantum measurements, so that the Haar distribution can be obtained from purely deterministic dynamics at the cost of conditioning on the bath measurements.  
An essential question lies in understanding under what assumptions conventional thermalization ($k=1$) implies deep thermalization ($k>1$)~\cite{ippoliti2022dynamical, wilming2022high} and whether additional information beyond conserved quantities are required to characterise the statistics of the projected ensemble. Focusing on non-interacting fermions, a recent study has shown that an ensemble of Gaussian states can satisfy Eq.~\eqref{eq:deepth} for $k=1$ but will exhibit deviations for higher moments, leading to the notion of \textit{generalised deep thermalisation}~\cite{PhysRevA.107.032215}.
Beyond the basic question of whether a quantum system thermalizes, it is important to understand whether there exists in general a hierarchy of time scales for the duration of the purely unitary evolution, known as design times $t_k$,  after which the measurements in $B$ gives a good $k$ design. For the solvable case of DUC, it has been shown that,
as long as the size $L_B$ of $B$ diverges and the region $A$ sits at one boundary,
for all $k$'s, $t_k = L_A$ with perfect accuracy, with $L_A$ the length of the region $A$. This feature is not expected to be generic, but rather to be a consequence of the strong chaotic nature of dual-unitary circuits and the fine-tuning that underlies them. In fact, the recent study~\cite{ippoliti2022solvable} introduced a solvable model based essentially on a simple random circuit consisting of three coarse-grained sites, representing region $A$, bath $B$, and the proxy $C$ sitting in between $A$ and $B$ themselves, respectively. The time evolution is performed via random unitary gates $2$--site~\cite{Nahum2017, Nahum2017a, vonKeyserlingk2017, doi:10.1146/annurev-conmatphys-031720-030658} so that $A$ and never $B$ are never interacting directly: the bottleneck mediating the interaction between region $A$ and bath $B$ provides a mechanism by which the time scales $t_k$ actually acquire a dependence on $k$, albeit only logarithmic $t_k = t_1 + O(\log k)$, valid for $k \ll d_A$.

In this work, we extend the study of deep thermalization to the case of spatially-extended random unitary circuits (RUC). RUC are constructed by evolving a spin chain using local gates drawn from some convenient distribution, e.g. the Haar measure over the unitary group, and acting between neighboring spins~\cite{doi:10.1146/annurev-conmatphys-031720-030658}. By leveraging the fact that each gate is drawn from the Haar ensemble, the average of many relevant quantities naturally requires an effective pairing of degrees of freedom in spacetime~\cite{Nahum2017}, eventually leading to a mapping onto a classical statistical mechanics model. The resulting classical models often become tractable at least in the large-dimensional limit of the local Hilbert space, $q \to \infty$. This approach paved the way toward understanding spectral statistics~\cite{bohigas1984characterization, cdc1, cdc2, cdc3, garratt2021dw}, which, beyond the late-time emergence of   the random matrix theory, displays 
early-time universal deviation~\cite{cdc2, chan2021manybody, Shivam_2023}, and operator spreading, a concept directly related to the butterfly effect and out-of-time order correlators~\cite{Nahum2017a, von_keyserlingk_operator_2018, PhysRevD.94.106002}. Furthermore, RUCs provided a solvable example of the growth of bipartite entanglement in chaotic circuits in the absence of quasiparticles~\cite{KimHuse}. More precisely, entanglement growth can be elegantly and effectively reinterpreted as the cost associated with a membrane (or domain wall) that separates different types of pairing in space-time~\cite{Nahum2017}. It has been argued that this formulation of entanglement in terms of a membrane is a general feature of chaotic systems, beyond the RUC in which it was actually derived~\cite{PhysRevX.10.031066, cdc1, rakovszky2019subballistic} and brought out the role of the Kardar-Parisi-Zhang (KPZ) equation~\cite{CORWIN2012} in describing its universal fluctuations in generic noisy dynamics~\cite{PhysRevB.99.174205, PhysRevB.98.184416}. More recently, RUC have been employed in the study of monitored dynamics, in situations in which unitary evolution is flanked by projective measures. This treatment has made it possible to conjecture the existence of measurement-induced phase transitions, which separate a volume law phase from an area law as the rate of projective measurements increases~\cite{Skinner_2019, Li_2018, Li_2019, 10.21468/SciPostPhys.7.2.024}.
In contrast to these cases, in this paper we focus on the aforementioned projected ensemble, where the simultaneous measurements of all sites in $B$ follows the purely unitary dynamics of the whole system, thus revealing a pure state in $A$. We stress that we are interested in the PE exclusively generated by measurements for a fixed realisation of the RUC performing the unitary dynamics. However, to take advantage of the Haar average of the circuit, we focus on the calculation of the frame potential $F^{(k)}_{\mathcal{E}}$, a quantity that measures the $k$-th moment of overlap between two random states in the ensemble $\mathcal{E}$. One can show (see below) that, among all ensembles, the Haar one provides the minimal frame potential, so that $F^{(k)}_{\rm PE} \geq F^{(k)}_{\rm Haar}$. Thus, because of such an exact inequality, the frame potential can be averaged over the realisation of the RUC, leading to an estimate of the distance from a $k$-design of each individual realisation. Remarkably, we found that the calculation of circuit-averaged frame potential leads to a model of statistical mechanics in the presence of domain-wall boundary conditions, similarly to what happens in the calculation of entanglement entropies. In this context, we can once again link the emergence of a $k$-design and deep thermalization to the existence of a membrane: at short times, the membrane starts as a vertical structure with a free end point exiting on the initial factorized state. However, as time progresses, it is less costly for the membrane to be pinned on the boundary, and the value of the frame potential saturates. In particular, for large $q$, the calculation of the frame potential in the projected ensemble can be exactly linked to the moments of the purity of the subregion $A$. Building on this membrane picture in generic models, we explore the saturation regime and estimate the behavior of the design times $t_k$. We discover that the logarithmic dependence, originally found in the toy model of \cite{ippoliti2022solvable}, is thus quite robust and has its origin in the exponential suppression of the unpinned configurations for the membrane at times $t \gtrsim L_A$.

The paper is organised as follows: In Sec.~\ref{sec:pse}, we provide the definitions of the frame potential, the Haar state ensemble, the projected ensemble and a description of the replica trick in this setting. In Sec.~\ref{sec:ruc}, we define a RUC with the brick-wall geometry, and express the frame potential of the RUC as a statistical mechanical problem. In Sec.~\ref{sec:fp_and_purity}, we evaluate the projected ensemble in the RUC by exploiting a connection between the frame potential and the purity in the large-dimensional limit of the local Hilbert space $q$. In Sec.~\ref{sec:mem}, we discuss the frame potential in the membrane picture at finite values of $q$. In Sec.~\ref{sec:design_time}, we study the saturation of the frame potential and show that the $k$-th design time displays a logarithmic dependence in $k$, in agreement with numerical simulations. Lastly, in Sec.~\ref{sec:non_local}, we discuss the notion of non-locality and long-range order which arise from the projected ensemble with modified boundary conditions.

\section{Ensembles of pure states\label{sec:pse}}
\subsection{Frame potential}
We start considering a generic ensemble of pure states defined by the set of pairs
\begin{equation}
 \mathcal{E} = \{ p(i), \ket{\Psi(i)}\}_{i} \;,
\end{equation}
where, with a slight abuse of notation, we include the possibility that the index $i$ runs on a continuous measure of states. The values $p(i)$ are the probabilities of picking the state $\ket{\Psi(i)}$ and are thus normalised $\sum_i p(i) = 1$. To fix the notation and for consistency with what we will use later, we assume that the states are chosen within the Hilbert space $\mathcal{H}_A$ of the region $A$, containing $L_A$ $q$-dit spins and thus have dimension $\HH_A = q^{L_A}$. The ensemble $\mathcal{E}$ can be seen as a possible unraveling of the density matrix 
\begin{equation}
\rho^{(1)}_\mathcal{E} := \sum_i p(i) \ket{\Psi(i)}\bra{\Psi(i)} \;.
\end{equation}
Of course, the same density matrix $\rho^{(1)}_{\mathcal{E}}$ leads to multiple possible unravelings~\cite{breuer2002theory}. A more refined characterization of the ensemble $\mathcal{E}$ is obtained by looking at the higher moments, or the \emph{$k$-replica density matrices}
\begin{equation}
    \rho^{(k)}_{\mathcal{E}} := \sum_{i} p(i) 
    \left( \ket{\Psi(i)} \bra{\Psi(i)} \right)^{\otimes k} \;,
\end{equation}
where $\Tr[\rho^{(k)}_{\mathcal{E}}] = 1$ is satisfied for all $k$'s.
We can measure the degree of ergodicity with which the ensemble covers the accessible Hilbert space by introducing frame potentials
\begin{equation}
    F^{(k)}_{\mathcal{E}} := \sum_{i,i'} p(i) p(i') |\braket{\Psi(i) | \Psi(i')}|^{2k} = \Tr[(\rho_{\mathcal{E}}^{(k)})^2] \;,
\end{equation}
which, can also be interpreted as the purity of the $k$-th density matrix. Alternatively, one can see $F^{(k)}_{\mathcal{E}}$ as the moments of the overlaps $|\braket{\Psi(i) | \Psi(i')}|$ between two random states sampled from $\mathcal{E}$.
\subsection{Haar state ensemble}
A yardstick of state ensemble is given by the uniformly distributed random states over the Hilbert space of the region $A$. This is given by the ensemble
\begin{equation}
    \eH = \{ \haarU \ket{\Psi} \; | \; \haarU \in \mbox{Haar}[\mathcal{H}_A] \} \;, 
\end{equation}
where $\ket{\Psi}$ is a 
 reference state and $\haarU$ is a Haar-distributed unitary matrix of size $\HH_A \times \HH_A$. The corresponding $k$-replica density matrices 
\begin{equation}
    \rhoH^{(k)}  := \int_{\ket{\Psi} \in \eH}  d \Psi \; \ket{\Psi} \bra{\Psi} ^{\otimes k} \;.
\end{equation}
can be explicitly evaluated~\cite{Collins_2022}
\begin{equation}
        \rho^{(k)}_{\rm H} = \frac{\sum_{\sigma \in S_k} \sigma}{\HH_A (\HH_A + 1)\ldots (\HH_A + k - 1)} \stackrel{\HH_A \gg 1}{\sim} \HH_A^{-k} \sum_{\sigma \in S_k} \sigma \;, 
\end{equation}
where $S_k$ represents the permutation group with $k$ elements and $\sigma$ acts on $\mathcal{H}_A^{\otimes k}$ by permuting the copies. From this, we deduce the expression for the frame potentials in the Haar ensemble as 
\begin{equation}
   \FH^{(k)} = \Tr[(\rhoH^{(k)})^2] = 
    \frac{k!}{\HH_A (\HH_A + 1)\ldots (\HH_A + k - 1)} 
    = \binom{\HH_A + k -1}{k}^{-1} \stackrel{\HH_A \gg 1}{\sim} \frac{k!}{\HH_A^k} \;.   \label{eq:Fkhaar}
\end{equation}
To compare how different a given ensemble is from the Haar state ensemble, we define a notion of distance using the $k$-th replica density matrix
\begin{equation}
\label{eq:distance}
    \Delta^{(k)} := 
    \frac{
    \dist{ \rho^{(k)}_{\mathcal{E}} - \rhoH^{(k)} }}
    {\dist{\rhoH^{(k)}}}\;, 
\end{equation}
where  $\dist{\cdot }$ is an appropriately chosen norm. For technical reason, we focus on the Frobenius norm
\be
|| M || = \sqrt{\Tr\left(M^\dag M \right)} \;.
\ee
With this definition, we can easily relate the distance to the frame potentials
\begin{equation}
\label{eq:DeltakF}
    [\Delta^{(k)}]^2 = \frac{F^{(k)}_{\mathcal{E}}}{F_{\rm H}^{(k)}} - 1  \;,
\end{equation}
where we used in the last equality
\begin{equation}
\Tr[ \rho^{(k)}_{\mathcal{E}}  \rhoH^{(k)}] = \frac{1}{{\HH_A (\HH_A + 1)\ldots (\HH_A + k - 1)}}\sum_{\sigma \in S_k} \Tr[\sigma \rho^{(k)}] = \FH^{(k)} \;,
\end{equation}
since $\Tr[\sigma \rho^{(k)}] = \Tr[\rho^{(k)}] =1 $ for any permutation $\sigma$, as $\rho^{(k)} $ is a replicated density matrix for $k$ identical copies and $\sigma \ket{\Psi}^{\otimes k} =  \ket{\Psi}^{\otimes k}$.
From Eq.~\eqref{eq:distance}, we see that $F^{(k)}_{\mathcal{E}} \geq \FH^{(k)}$ and the equality is reached when the ensemble $\mathcal{E}$ is indistinguishable from the Haar one up to the moment $k$ and is thus a $k$-design.

\subsection{Projected state ensemble} 
The projected ensemble (PE) is based on the following idea: Given a quantum state $\ket{\Psi}$, belonging to a Hilbert space which can be partitioned into two complementary regions $A$ and $B$, we can measure a complete set of local observables on $B$, and look at the resulting pure state on $A$ (see for example Fig.~\ref{fig:projected_state_ensemble}). More specifically, 
the ensemble of projected states is given as a set of pairs of probabilities and states
\begin{equation}
    \EPE := \{ p(\aaa), \ket{\Psi_A(\aaa)}\}_{\aaa} \;,
\end{equation}
where  $\aaa = \{a_j \}_{j = 1}^{L_B}$ labels 
the outcomes of all measurements in the region $B$ of size $L_B$, in a fixed defined basis with $a_j \in \{1,\ldots, q\}$. 
The statistical properties of the ensemble of pure states $\EPE$ provide a refined characterization of the quantum correlations in the original state $\ket{\Psi}$. In our case, we are interested in a situation where the state $\ket{\Psi}$ has been generated dynamically by $\UU(t)$, the time evolution operator governing the system's dynamics. Denoting as $\ket{\Psi(t)} = \UU(t) \ket{\Psi(0)}$ and with $\Pi_B(\aaa)$ the projector of the Hilbert space of $B$ onto the observed outcomes $\aaa$, we obtain from the Born's rule that
\begin{equation}
\label{eq:projpsip}
 \ket{\Psi_A(\aaa)} :=  \frac{\Pi_B(\aaa) \ket{\Psi(t)}}{p(\aaa)^{1/2}} \;, \qquad  p(\aaa)  := \braket{\Psi(t)|\Pi_B(\aaa) |\Psi(t)} \;.    
\end{equation}

\subsection{Replica trick}
It is useful to define the time-evolved and measured state without the normalization as
\begin{equation}
    \ket{\tilde{\Psi}_A(\aaa)} :=  \Pi_B(\aaa) \UU(t) \ket{\Psi}  \;.
\end{equation}
Then, the frame potential can be written as
\begin{equation}
\label{eq:frameU}
    F^{(k)}_{U(t)} =  \sum_{\aaa, \aaa'} \braket{\tilde{\Psi}_A(\aaa) | \tilde{\Psi}_A(\aaa) }^{1-k} \braket{\tilde{\Psi}_A(\aaa') | \tilde{\Psi}_A(\aaa') }^{1-k} |\braket{\tilde{\Psi}_A(\aaa) | \tilde{\Psi}_A(\aaa') }|^{2k} \;,
\end{equation}
where the factor of $\braket{\tilde{\Psi}_A(\aaa) | \tilde{\Psi}_A(\aaa) }^{1-k}$ (and similarly for $\aaa \to \aaa'$) originates from the normalization of the state $\ket{\tilde{\Psi}_A(\aaa)}$ after projection and the Born's probabilities $p(\aaa) = \braket{\tilde{\Psi}_A(\aaa) | \tilde{\Psi}_A(\aaa) }$.
The main difficulty in evaluating this expression is the presence of negative integer exponents $1-k$ for any $k>1$. To avoid it, we make use of the replica trick and set $n = 1-k$, which we treat temporarily as a positive integer~\cite{Jian_2020, PhysRevB.100.134203, altman2019, nahum2023renormalization, giachetti2023elusive}.  This leads to
\begin{equation}\label{eq:fp_replica}
    F^{(k,n)}_{U(t)} :=  \sum_{\aaa, \aaa'} \braket{\tilde{\Psi}_A(\aaa) | \tilde{\Psi}_A(\aaa) }^{n} \braket{\tilde{\Psi}_A(\aaa') | \tilde{\Psi}_A(\aaa') }^{n}  |\braket{\tilde{\Psi}_A(\aaa) | \tilde{\Psi}_A(\aaa') }|^{2k} \;.
\end{equation}
Then, if the parameter $n$ appears in a way that can be analytically continued, we can obtain the result from the limit
\begin{equation}
\label{eq:replicalim}
    \lim_{n \to 1-k} F^{(k,n)}_{U(t)} = F^{(k)}_{U(t)} \;.
\end{equation}
Since in Eq.~\eqref{eq:fp_replica} the exponent $n$ is simply associated with the norm of each state, in what follows, we will use the nomenclature of \textit{spectators} for the $n$ replicas, as opposed to the remaining $k$ whose overlap is calculated. 

\section{Random unitary circuits}\label{sec:ruc}
\subsection{Model}
\begin{figure}[h]
\begin{center}
\includegraphics[width=0.6\textwidth]{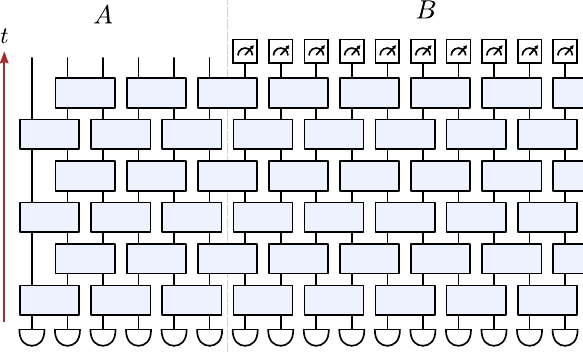}
\end{center}
\caption{ Diagrammatic illustration of the projected state ensemble for a random quantum circuit as a model of many-body quantum chaotic systems. 
Space and time run horizontally and vertically, respectively. The system is prepared in a product state, evolved under the brick wall model, and is projectively measured in region $B$. 
\label{fig:projected_state_ensemble}}
\end{figure}
We now consider the projected ensemble starting from a state which is generated by a realisation of the brick-wall random circuit model (BWM), as illustrated in Fig.~\ref{fig:projected_state_ensemble}. The BWM is a random quantum circuit with the brick-wall geometry, defined as 
\begin{align}
\UU(t) &:= \prod_{\tau=1}^t \, \UU_{0}(\tau)
\,
\UU_{1}(\tau)\;, \label{eq:UtU0U1}
\\
U_{a}(\tau) &:= \bigotimes_{j \in 2\mathbb{Z} +a} \uu_{j, j+1}(\tau)  \;,
\end{align}
where the local dimension of the Hilbert space is taken to be $q$, and $u_{j, j+1}(\tau)$ are drawn independently according to the Haar measure $\mbox{Haar}[\mathbb{C}^{q^2}]$ on the Hilbert space of two sites.
In the following, we will use $\overline{(\dots )}$ to denote ensemble average over the realizations of the BWM.
In this context, the state $\ket{\Psi(t)}$ in \eqref{eq:projpsip}
is chosen as $\ket{\Psi_\UU(t)} \equiv \UU(t) \ket{\Psi_0}$, where $\UU(t)$ is a specific realization of the BWM circuit. In order to investigate the scrambling induced by the quantum dynamics, we assume that $\ket{\Psi_0}$ is any factorized product state (the specific choice is irrelevant because of the Haar invariance of the applied gates).
It is our focus to evaluate how far is the projected ensemble from the Haar one for a given realisation of the BWM. However, to take advantage of the random matrix formalism, we focus on the circuit-averaged frame potentials
\begin{equation}
    F^{(k)}_{\rm BWM} := \overline{F^{(k)}_{U(t)}} \;.
\end{equation}
In fact, because of the exact inequality $F_{U(t)}^{(k)} \geq F_{H}^{(k)}$, convergence to Haar of the average frame potential $F^{(k)}$ places a limit on the probability that each realization of the BWM has a significant deviation.
In more detail, Markov's inequality guarantees that
\begin{equation}
    \mbox{Prob}\left(F^{(k)}_{U(t)} - F_{H}^{(k)} > \epsilon \right) \leq \frac{\overline{F^{(k)}_{\rm BWM} - F^{(k)}_H}}{\epsilon} \;.
\end{equation}
where the probability is computed over the realisations of the BWM.

\subsection{Set-up \label{sec:setup}}
\begin{figure}[h]
\begin{center}
\includegraphics[width=0.7\textwidth]{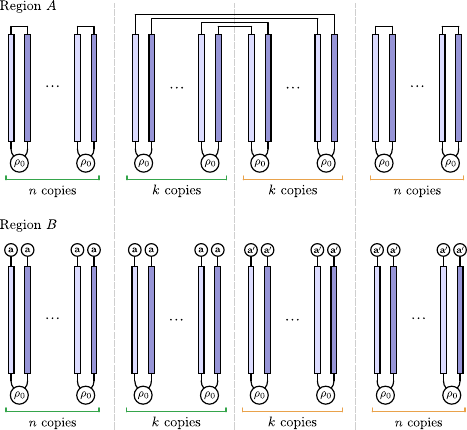}
\caption{Diagrammatic representation of  $F^{(k,n)}$  of sites in region A (top) and region B (bottom), where time runs vertically  and space runs perpendicular to the page. 
$\uu$ and $\uu^\dagger$ are represented by light and dark blue gates respectively. 
For a given site in space, we label each $\uu$ (and $\uu^\dagger$) from $1$ to $2n+2k$, from left to right in this diagram. 
\label{fig:frame_rho}}
\end{center}
\end{figure}

Diagrammatically, we represent the frame potential in Eq.~\eqref{eq:fp_replica} as in Fig.~\ref{fig:frame_rho}. 
Considering both the terms involving $\aaa$ and $\aaa'$, we have a total of $2n + 2k$ bras 
and kets, 
each bringing a copy of the circuit $U(t)$ and $U^\dag(t)$ respectively. 
This means that each gate $u_{j,j+1}(\tau)$ and its complex conjugate $u^\dag_{j,j+1}(\tau)$, for each pair of sites $(j,j+1)$ and time $\tau$, will appear $m = 2n+2k$ times.
Because the gates are chosen with Haar distribution, averaging several copies is equivalent to pairing the unitaries $u$ with their complex conjugates $u^\dag$~\cite{Collins_2022}. To express the result quantitatively, it is useful to vectorize the density matrix
\begin{equation}
    \rho^{(m)} \to \kket{\rho^{(m)}}
\end{equation}
where $\kket{\rho^{(m}}$ is now a state in a doubled Hilbert space.
Consider the action of the $m$ copies of unitaries, $\uu$ and $\uu^\dagger$, which act on sites $j$ and $j+1$. Upon averaging, the unitaries become a sum over two sets of permutations, according to the identity~\cite{Brouwer1996,8178732}
\begin{equation}
\label{eq:uuwg}   \overline{\underbrace{\uu \otimes \uu \ldots \otimes \uu}_{\mbox{$m$ times}} \otimes \underbrace{\uu^\dag \otimes \uu^\dag \ldots \otimes \uu^\dag}_{\mbox{$m$ times}}} = \sum_{\sigma, \tau \in S_m} \Wg(\sigma \tau^{-1}; q^2) \kket{\sigma}_j\kket{\sigma}_{j+1}\prescript{}{j}{\bbra{\tau}}\prescript{}{j+1}{\bbra{\tau}}  \;,
\end{equation}
where to shorten the notation, we omit the site indices $u_{j,j+1}$ on the left hand side.
Here, we introduce a ket / bra notation for permutations in the replica space. The notation $\kket{\sigma}_j$ denotes a particular pairing of replicas for the site $j$: the components of a permutation $\sigma \in S_m$ over the replicas $2m$ ($m$ bras and $m$ kets) of each site are
\begin{equation} 
\label{eq:expsigma}
\bbraket{a_1, \bar a_1, \ldots, a_m, \bar a_m \doublev \sigma} = \prod_{i=1}^m \delta_{a_i, \bar a_{\sigma_i}}  \;,
\end{equation}
where we denoted as $\{a_\ell,  \bar{a}_\ell\}_{\ell=1}^m \in \{1,\ldots, q\}$ the components of each replica in the orbital basis. From this it follows that different permutations are not orthogonal, but have the scalar product.
\begin{equation}
\label{eq:scalarprod}
\bbraket{\sigma\doublev\tau} = q^{N_c(\sigma \tau^{-1})} = q^{m - d(\sigma, \tau)
}\;,
\end{equation}
where $N_c(\sigma)$ counts the number of cycles in the permutation $\sigma \in S_m$ and $d(\sigma, \tau) = m - N_c(\sigma \tau^{-1})$ is the transposition distance between $\sigma$ and $\tau$. 
Using the two-row notation for permutations,  $\sigma: \ell \to \sigma(\ell)$
\begin{equation}
  \sigma =
  \begin{pmatrix}
    1 & \cdots & m \\
    \sigma(1) & \cdots  & \sigma(m)
    \end{pmatrix} \;,   
\end{equation}
we define some useful permutations: 
\begin{itemize}
\item the identity permutation
\be
  \id_{m} =  
  \begin{pmatrix}
    1 & \cdots & m \\
    1 & \cdots  & m
    \end{pmatrix} \;,
\ee
\item the inversion permutation
\be
\label{eq:pswap}
  \imath_{2m} =  
  \begin{pmatrix}
    1 & \cdots & m & m+1 & \cdots  & 2m\\
    2m & \cdots  & m+1 & m & \cdots & 1
    \end{pmatrix} \;.
\ee
\item the translation permutation
\be
  \tau_{m} =  
  \begin{pmatrix}
    1 & 2 &\cdots & m \\
    m & 1 &\cdots  & m-1
    \end{pmatrix} \;,
\ee
\end{itemize}
Furthermore, given $\sigma\in S_m$ and $\sigma' \in S_{m'}$, we define the operation $(\cdot , \cdot): S_m \times S_{m' } \to S_{m+m'}$ by
\be
\label{eq:embed}
  (\sigma,\sigma') =  
  \begin{pmatrix}
    1 & \cdots & m & m+1 & \cdots  & m+m'\\
    \sigma(1) & \cdots & \sigma(m) & m+ \sigma'(1) & \cdots  & m+ \sigma'(m')
    \end{pmatrix} \;.
\ee
which can be extended to more than two permutations by associativity, i.e. $(\sigma, \sigma', \sigma'') := (\sigma, (\sigma', \sigma''))$.
%
%
With these notations, we can now express the boundary conditions on the top boundary for both sites in the region $A$ and $B$ (see Fig.~\ref{fig:frame_rho}). 
To arrange the $2n+2k$ replicas, it is convenient to put the first (last) $n$ as spectators relative to the set of measures $\aaa$ ($\aaa'$). Then, in region $A$ (Fig.~\ref{fig:frame_rho} top), the boundary state has the form 
\be
\label{eq:muA}
\bbra{A} = \bbra{\mu_A}\;,  \qquad  \mu_A = (\id_n, \imath_{2k} , \id_n) \in S_{2n + 2k}\;,
\ee
corresponding to the identity in both sets of the spectator replicas $\mathbb{1}_n$ and to the inversion~\eqref{eq:pswap} for the middle. 
For the sites in $B$, we instead have
\begin{equation}
\bbra{B} = \sum_{a, a' = 1}^q  \bbra{
\underbrace{a,a,\ldots, a}_{\mbox{\scriptsize $n+k$ times}} , \underbrace{a',a',\ldots, a'}_{\mbox{\scriptsize $n+k$ times}} 
}\;.
\end{equation}
From this representation and \eqref{eq:expsigma}, we can easily express the overlap between the state $\bbra{B}$ and any permutation $\sigma$. Indeed, the overlap will favor permutations that do not mix the first/last $n+k$ replicas. More precisely, one has
\begin{equation}
\label{eq:Bsigma}
\bbraket{B | \sigma} = \sum_{a,a'} [\mathbf{1}_F(\sigma)  + (1 - \mathbf{1}_F(\sigma)) \delta_{a,a'}] = 
\mathbf{1}_F(\sigma)(q^2 - q) + q \;,
\end{equation}
where we introduced $\mathbf{1}_F(\sigma)$ as the characteristic function of factorised permutations, i.e.
$\mathbf{1}_F(\sigma) = 1$ if $\sigma = (\sigma_1, \sigma_2)$ with $\sigma_{1,2} \in S_{n+k}$ and zero otherwise.

From the states $\bbra{A}$ and $\bbra{B}$ for the sites $j \in A$ and $j \in B$, respectively, we define the top state $\bbra{\mathcal{F}}$ at the final time as a tensor product of states in space
\begin{equation}
\label{eq:Fdef}
    \bbra{\mathcal{F}} = \prescript{}{1}{\bbra{A}} \ldots \prescript{}{L_A}{\bbra{A}} \prescript{}{L_A + 1}{\bbra{B}} \ldots \prescript{}{L}{\bbra{B}} \;.
\end{equation}
which indeed corresponds to a domain-wall boundary condition.
On the contrary, the initial state is a tensor product of the same states across space, as given by
\begin{equation}
    \kket{\mathcal{I}} = \kket{0}_1 \ldots \kket{0}_L \;,
\end{equation}
where  $\kket{0}$  is defined in the replicated space by $\bbraket{\sigma | 0} = 1$.
Using Eq.~\eqref{eq:uuwg} for each group of replicated $2$-site gates,  we obtain a transfer matrix $\mathcal{T}$ for each bi-layer in the circuit, which inherits the brick-wall structure. This leads to the expression
\begin{equation}\label{eq:fkn}
    F^{(k,n)}_{\rm BWM} = \bbraket{\mathcal{F} | \mathcal{T}^t | \mathcal{I}} \;, 
\end{equation}
which can be seen as a classical statistical mechanics model on a strip where the local degrees of freedom are permutations and the Boltzmann's weights are controlled by the Weingarten's functions \eqref{eq:uuwg} and the scalar products \eqref{eq:scalarprod}. 
The weights have an intricate structure and can be negative, which complicates the analytical calculation needed for the replica limit $n \to 1-k$. For this reason, we proceed by considering the large-$q$ limit. Additionally, with the notations introduced, in Appendix \ref{app:pe_haar} we provide the evaluation of the projected ensemble for the Haar state ensemble as an instructive example.

\section{Mapping between frame potential and purity at large $q$}\label{sec:fp_and_purity}

\begin{figure}[h]
\begin{center}
\includegraphics[width=0.6\textwidth]{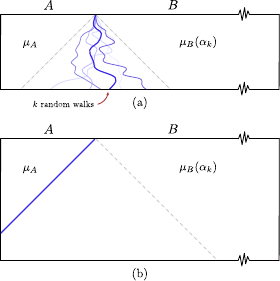}
\caption{
Illustration of the frame potential $F^{(k,n)}$ where time and space run vertically and horizontally respectively, and the system size of region $A$ is much smaller than the one of region $B$, i.e. $L_A \ll L_B$.
(a) For $t \leq L_A/2$, the frame potential is a sum of configurations with $k$ uncorrelated random walks of domain walls.   
(b) For $t > L_A/2$, the frame potential is the $k$ copies of the shortest walks from the boundary of $A$ and $B$, towards the opposite boundary of $A$. Note that, at large local Hilbert space size $q$, $\mu_A$ is a fixed permutation in $S_{2n + 2k}$, while $\mu_B(\alpha_k)$ can assume $k!$ values according to some permutation $\alpha_k \in S_k$, see Eq.~\eqref{eq:permutation_alpha}.
\label{fig:membrane}}
\end{center}
\end{figure}

In the large-$q$ limit, Eq.~\eqref{eq:uuwg} can be hugely simplified using the expansion
\begin{equation} \label{eq:wg_largeq}
    \Wg(\sigma \tau^{-1}; q^2) \stackrel{q \gg 1}{=} 
    \frac{\delta_{\sigma, \tau}}{q^{2m}} [1 + o(1)]\;.
\end{equation}
Additionally, at large $q$, the scalar product \eqref{eq:scalarprod} also forces $\sigma = \tau$, as the identity $\mathbb{1}_{n+k} = \sigma \tau^{-1}$ maximises the number of cycles.
This means that the application of the transfer matrix $\mathcal{T}$ from the boundary state $\kket{\mathcal{F}}$ tends to maintain the same permutation whenever possible.
In particular, deep inside region $A$, the top boundary condition fixes the permutation to be $\mu_A$ in Eq.~\eqref{eq:muA}. 
 Deep in the region $B$, according to Eq.~\eqref{eq:Bsigma}, the boundary condition at large $q$ gives equal weights to all factorized permutations $\mu_B = ( \sigma_{n+k}, \sigma'_{n+k} ) \in S_{n+k} \times S_{n+k}$.
%
More precisely, this implies that the contractions in $A$, outside of the light cone of the boundary between $A$ and $B$, take the following permutations $\mu_A = ( \id_n, \imath_{2k}, \id_n ) \in S_{2n+2k}$. Instead, in the bulk of the region $B$, at large $q \to \infty$, Eq.~\eqref{eq:Bsigma} favours any factorized permutations, i.e.  
$\mu_B = ( \sigma_{n+k}, \sigma'_{n+k} ) \in S_{n+k} \times S_{n+k}$, as illustrated in Fig.~\ref{fig:membrane}. 
Moreover, any spatial variation of the pair $( \sigma_{n+k}, \sigma'_{n+k} )$ within the region $B$ would lead to an additional interface and thus an additional suppression factor when $q \to \infty$. We are thus left with evaluating the cost of the interface between $\mu_A$ and $\mu_B$. 
First, we observe that we can use the residual freedom in the choice of $\mu_B$ to reduce the cost of the interface. As it follows from the structure of the scalar product \eqref{eq:scalarprod} and further explained in \cite{PhysRevB.99.174205}, the cost of the interface is directly associated with the transposition distance between permutations $\mu_A$ and $\mu_B$ across it. Because $\mu_A$ acts as the identity on the first/last $n$ replicas, the optimal choice must be of the form $\sigma_{n+k} = (\id_n, \alpha_k)$ and $\sigma'_{n+k} = (\alpha'_k, \id_n)$ with $\alpha_k, \alpha'_k \in S_k$.
Furthermore, to optimize the transposition distance
$ d(\imath_{2k}, (\alpha_k, \alpha'_k)) $, $\alpha'_k$ is completely fixed by $\alpha_k$ via the relation
\be\label{eq:permutation_alpha}
\ba
(\mathbb{1}_k, \alpha_k') = \imath_{2k} \cdot (\alpha_k^{-1}, \mathbb{1}) \cdot \imath_{2k} \;,
\ea
\ee
i.e. $\alpha_k'$ is obtained by inverting $\alpha_k$ and performing reflections with $\imath_{2k}$.
 Then, $(\alpha_k, \alpha_k') = (\alpha_k, \mathbb{1}_k) \cdot (\mathbb{1}_k, \alpha_k') = (\alpha_k, \mathbb{1}_k) \cdot \imath_{2k} \cdot (\alpha_k^{-1}, \mathbb{1}) \cdot \imath_{2k}$. 
 Therefore, we have $d(\imath_{2k}, (\alpha_k, \alpha'_k)) = d((\alpha_k, \alpha_k') \cdot \imath_{2k}^{-1}, \id_{2k}) =d((\alpha_k, \mathbb{1}_k) \cdot  \imath_{2k} \cdot  (\alpha_k^{-1}, \mathbb{1}_k), \id_{2k})= d(\imath_{2k}, \id_{2k}) = k$ for all $\alpha_k$. 
 Additionally, we can explicitly decompose
 \begin{equation}
 \label{eq:decomptransp}
     (\alpha_k, \alpha_k') \cdot \imath_{2k} = \prod_{j=1}^k \tau_{\alpha_j,2k-j+1} \;, 
 \end{equation}
where $\tau_{ij}$ is the transposition permutation exchanging $i$ with $j$. Therefore, to highlight the fact that $\mu_B$ depends on $\alpha_k \in S_k$, we write $\mu_B(\alpha_k)$.

At this point, it is useful to take a small detour by recapitulating the calculation of the purity in the same context of BWM. Consider the definition of the purity for the subsystem $A$ as the trace of the squared reduced density matrix of the subsystem $A$
\begin{equation}
\label{eq:puritydef}
    \mathcal{P} := \Tr_A \left[\Tr_B [ 
U(t) \ket{\Psi} \bra{\Psi} U^\dag(t)]^2\right]=
 \sum_{\aaa, \aaa'}  |\braket{\tilde{\Psi}_A(\aaa) | \tilde{\Psi}_A(\aaa') }|^{2} = F_{U(t)}^{(1)} \;,
\end{equation}
where in the last equality we used the definition \eqref{eq:frameU}. Thus, if we consider the $k$-th moment under the circuit average, we can once again express it in the formalism of \ref{sec:setup} as $\overline{\mathcal{P}^k} = 
 \bbraket{\mathcal{F} | \mathcal{T}^t | \mathcal{I}}$, with $\bbra{\mathcal{F}}$ as in Eq.~\eqref{eq:Fdef} and 
\begin{equation}
\label{eq:transpPur}
\bbra{A_\mathcal{P}} := \bbra{\tau_{1,2}\tau_{3,4}\ldots \tau_{2k-1,2k}} \;, \qquad \bbra{B_\mathcal{P}} :=\bbra{\id_{2k}}   \;.
\end{equation}
We see the analogy emerging at large $q$ between the calculation of the frame potential $\overline{F^{(k,n)}_{U(t)}}$ needed for the projected ensemble and $\overline{(F^{(1)}_{U(t)})^k}$ which instead leads to the moments of the purity. In particular, an intrinsic property of both computations is that one is left with a domain-wall boundary condition with states in $A$ and $B$ differing by a product of $k$ commuting transpositions. In this case, in leading order in $q \to \infty$, since the permutations at the top boundary of frame potential factorize in Fig.~\ref{fig:frame_rho}, the $k$ domain walls behave as independent random walks, see Fig.~\ref{fig:membrane}. Following the derivation in \cite{cdc1}, we write
\begin{align}\label{eq:fp_purity_k}
F^{(k,n)}_{\rm BWM} &\stackrel{q \gg 1}{=}
\frac{k! }{q^{2(n+k-1)L_B}} \overline{\mathcal{P}^k}  \\
& =
\frac{k!}{q^{2(n+k-1)L_B}}
\left\{
\begin{aligned}
&   4^{kt} q^{-2kt}   \quad   & t \leq L_A/2 \;, \\
&   q^{-k L_A }  \quad  & t >L_A/2 \;.
\end{aligned}
\right.
\label{eq:fp_purity_k_res}
\end{align}
The factor of $k!$ is due to the sum over $\alpha_k \in S_k$, while the factor $q^{2(n+k-1)L_B}$ results from the prefactor in Eq.~\eqref{eq:wg_largeq} with $m = 2(n+k)$, combined with the $q^{2L_B}$ coming from the sum over $\aaa, \aaa'$. 
For $t \leq L_A/2$, the factor of $4^{kt}$ is associated to the entropy of the walks, i.e. it counts the number of possible random walks of domain walls. For clarity we recall that, with our conventions, a single time step $\Delta t = 1$ corresponds to the application of the two layers of the circuit (see Eq.~\eqref{eq:UtU0U1})
and thus two steps for the random walks. 
In a coarse-grained picture, it is useful to interpret the factor $4^{kt} q^{-2kt}$ as the cost associated to the line tension of the $k$ paths coming out from the bottom (see Fig.~\ref{fig:membrane} a).
For $t > L_A/2$, there is one type of paths dominating the sum for $q\to\infty$, that is, the walk taking $L_A$ steps from the partition between $A$ and $B$, to the boundary of $A$ (see Fig.~\ref{fig:membrane} b). 
Since the RHS has a simple dependence on $n$, the replica limit, $n \to 1-k$, is easily taken in the large-$q$ limit (with order of limits where $q$ is sent to infinity first) and leads to the expression
\begin{equation}\label{eq:fp_replica_res}
F^{(k)}_{\rm BWM} \stackrel{q\gg1}{=}
k! \, \overline{\mathcal{P}^k} = \left\{
\ba
&   k! 4^{kt} q^{-2kt}   \quad   & t \leq L_A/2 \;, \\
&   k! q^{-k L_A }  \quad  &t >L_A/2 \;.
\ea
\right.
\end{equation}
This result shows that in the leading order at large $q$ the saturation of the frame potentials occurs with a sharp transition for all design times $t_k = L_A/2$ irrespectively of $k$. As we will see below, the sharp transition at $t=O(L_A)$ will smoothen out at finite values of $q$ due to comparable contributions from configurations where walks exit at the bottom or left boundaries (see Fig.~\ref{fig:membrane}). A similar behavior to Eq.~\eqref{eq:fp_replica_res} has been observed in dual-unitary circuits~\cite{ho2022exact, ippoliti2022dynamical, claeys2022emergent} and is thus generally expected in extremely chaotic quantum many-body systems.

\section{Membrane picture at finite $q$}\label{sec:mem}
At finite $q$, we do not expect a quantitative agreement with Eq.~\eqref{eq:fp_replica_res} 
but a qualitative behavior should be captured by the domain-wall description. Let us first discuss the purity~\eqref{eq:puritydef}, or rather, of the second Renyi entropy $\mathcal{S}_2$, defined by the relation.
\begin{equation}
    e^{- \mathcal{S}_2} = \mathcal{P} \;, 
\end{equation}
summarising the results of \cite{PhysRevB.99.174205}. At short times $t$ with respect to the interval length $L_A$, the Renyi entropies are generally expected to grow linearly in time, a manifestation of the scrambling of the initial local information into non-local degrees of freedom~\cite{PhysRevLett.111.127205}. 
At large times, the dynamics is expected to locally thermalize. Since the BWM does not have conserved quantities, the behavior of finite subsystems at large times is captured by the infinite-temperature ensemble, with the volume law
\begin{equation}
\label{eq:puritysat}
    \mathcal{P} \stackrel{t \gg L_A}{\sim} e^{- s_{\rm eq} L_A} \;, \qquad s_{\rm eq} = \ln q \;.
\end{equation}
Corrections to \eqref{eq:puritysat} are subleading in volume in agreement with the predictions of random matrices~\cite{PhysRevLett.71.1291, Nadal2011, Nahum2017, Piroli2020, blake2023page}.
Both regimes can be understood in terms of the domain-wall dynamics introduced in the previous section, where Fig.~\ref{fig:membrane} retains its validity with some modifications. The moments $\overline{\mathcal{P}^k}$ can be understood in terms of $k$ domain walls, each formally represented by a transposition in the permutation group $S_{2k}$, see Eq.~\eqref{eq:transpPur}. Indeed, 
an ultimate consequence of the unitarity of the dynamics, which holds for arbitrary values of $q$,  is that the rules induced by \eqref{eq:uuwg} prevent the creation or annihilation of domain walls~\cite{PhysRevB.99.174205}. However, at finite $q$, the paths are no longer independent and can recombine, giving rise to an effective interaction. At short times, one can neglect the boundaries and the linear growth of Renyi entropies can once again be associated with the line tension of the domain walls exiting the bottom. For $k=1$, we have to deal with a single domain wall and the expression 
of the average purity before saturation is exactly computable for any $q$~\cite{Nahum2017, Nahum2017a, PhysRevX.8.021013}
\begin{equation}
\label{eq:purityexact}
    \overline{\mathcal{P}} = \left(\frac{2q}{1 + q^2}\right)^{2t} = e^{- \tilde{v}_2 s_{\rm eq} t} \;, \qquad \tilde{v}_2 = \frac{2\log(\frac{q^2+1}{2 q})}{\log q} \simeq 2\left(1 - \frac{\log 2}{\log q}\right)
\end{equation}
where $\tilde{v}_2$ is known as the purity speed, and again $s_{\mathrm{eq}}= \ln q$. When dealing with $k>1$, an exact treatment is not possible in general due to the interactions between the domain walls. However, the first correction at large $q$ is a mild attraction of order $O(q^{-4})$. The consequences of the interaction between the domain walls in moments $\overline{\mathcal{P}^k}$ 
are a manifestation of the fluctuations of the second Renyi entropy. Quantitatively, one can interpret $\mathcal{S}_2$ as the free energy of a single path in a quenched disorder~\cite{KARDAR1987582}: following~\cite{PhysRevB.99.174205}, this leads to
$     \mathcal{S}_2 = s_{\rm eq} (v_2 t + B_2 t^{1/3} \chi) $ in which the growth rate $v_2 = \tilde{v}_2 ( 1 + O(1/(q^8 \ln q)))$ is dressed and circuit randomness induces universal fluctuations on the $t^{1/3}$ scale with $B_2 = O(q^{-8/3})$ a non-universal coefficient vanishing at large $q$. The random variable $\chi$ follows the characteristic Tracy-Widom distribution $F_1$~\cite{Calabrese_2010, Dotsenko_2010, PhysRevLett.104.230602} induced by the Kardar-Parisi-Zhang (KPZ) equation~\cite{PhysRevLett.56.889}.
We stress that although the fluctuations $O(t^{1/3})$ in the second Renyi entropy $\mathcal{S}_2$ are
subleading at large times, they manifest themselves in the moments of the purity $\overline{P^k}$, whose decay rate acquires a non-trivial dependence in $k$.

\begin{figure}[h]
        \includegraphics[width=0.49\textwidth]{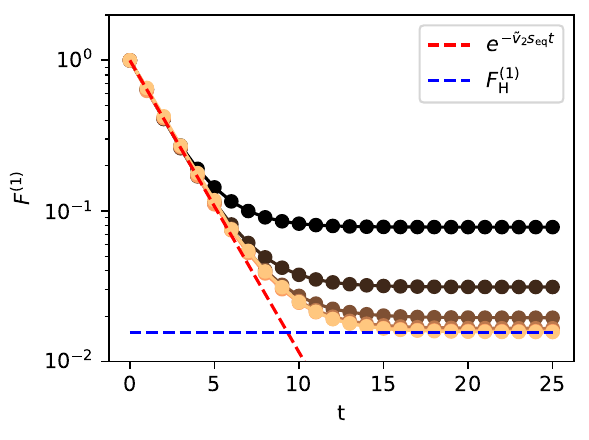}
    \hfill 
        \includegraphics[width=0.49\textwidth]{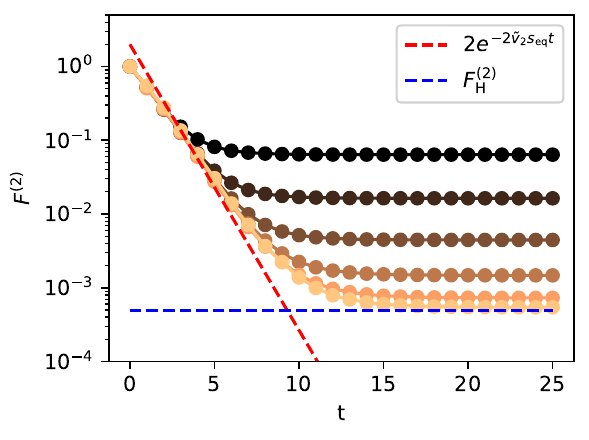}
    \hfill 
        \includegraphics[width=0.49\textwidth]{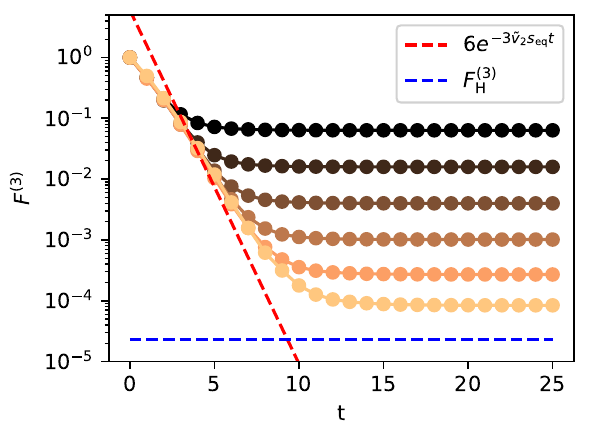}
    \caption{Numerical simulation of the frame potential $F^{(k)}_{\rm BWM}$ for $q=2$ and $L_A = 6$ with $k = 1,2,3$ and $L_B = 4,6,\ldots,16$ (black to orange). Comparison with the pre-saturation regime of Eq.~\eqref{eq:Fkmembrane} is shown in red, while the Haar saturation value is plotted in blue. Case $k=1$ is included as a benchmark, since the short-time decrease is exactly given by the average purity~\eqref{eq:purityexact}. Note that the non-interacting approximation overestimates the slope and the error increases with $k$: a numerical fit gives the estimates $v^{(k)} = k \tilde{v}_2 - \delta v^{(k)}$, with $\delta v^{(2)} \sim 0.7$ and $\delta v^{(3)} \sim 0.4$.}
    \label{fig:comparison_Fk}
\end{figure}

A similar picture can be applied to the behavior of the frame potentials $F^{(k)}_{\rm BWM}$, where, as we saw in the previous section, one has to deal with $k$ commuting domain walls. One thus expects
\begin{equation}
\label{eq:Fkmembrane}
    F^{(k)}_{\rm BWM} =  \begin{cases}
        k! e^{-v^{(k)} s_{\rm eq} t} \;, & t \ll L_A \;, \\
        F^{(k)}_{\rm H} \;, & t \gg L_A  \;.
    \end{cases}
\end{equation}
At large times $v_2 t \gg L_A$, the saturation $F_{\rm BWM}^{(k)} \to F_{\rm H}^{(k)} \sim k! e^{-k s_{\rm eq} L_A}$ is the consequence of all $k$ domain walls exiting the left boundary as in Fig.~\ref{fig:membrane}b. The decay rates $v^{(k)}$ result from the interaction of the walls of the $k$ domain, and we expect that again the $k$ dependence of $v^{(k)}$ is the result of a fluctuating interface, as described by the KPZ equation. However, as discussed in~\cite{PhysRevB.99.174205}, the crossover length $\ell_d = O(q^4)$ is rather long even at $q = 2$. In our units, the diffusion constant $D = 1/2$ and for times $t \lesssim \ell_d^2$, the $k$ domain walls behave as independent random walks, as the interactions between them are not yet relevant. Consequently, for relatively large values of $L_A$, saturation in Eq.~\eqref{eq:Fkmembrane} occurs well before the effect of interaction plays any role, and 
the sum over paths involved in the frame potential factorises as $k$ independent paths, thus implying that
we can approximate $v^{(k)} \sim k \tilde{v}_2$. A quantitative comparison between Eq.~\eqref{eq:Fkmembrane} and numerical simulations at $q=2$ is shown in Fig.~\ref{fig:comparison_Fk}. At the accessible system sizes, it is difficult to accurately estimate the slope $v^{(k)}$ and its dependence on $k$, but we observe that $v^{(k)} < k \tilde v_2$ and the difference increases with $k$. This is consistent with the effect of attractive interactions between replicas~\cite{PhysRevB.99.174205}. In other words, $F^{(k)}_{BWM}$ is larger at finite $q$ because multiple domain walls correlate by following the same favourable path. This gives an additional hint about the emergence of KPZ physics also for this quantity.

\section{Saturation and design times}\label{sec:design_time}
A finite $q$, not only the speed $v^{(k)}$ values in Eq.~\eqref{eq:Fkmembrane} are modified, but also the saturation at $t \approx L_A$ is rounded with respect to the sharp step in Eq.~\eqref{eq:fp_replica_res}. As we saw earlier, in calculating the frame potential using the transfer matrix, we can reduce to summing over the configurations of $k$ paths. Moreover, in practice for the accessible values of $L_A$, we can neglect the interactions between the paths. Therefore, near the saturation that occurs at times $t = O(L_A)$, the relevant configurations for each path are reduced to the one that ends at the left or the bottom boundary. For instance, for $k = 1$, one has~using~\cite{PhysRevB.99.174205}
\begin{equation}
\label{eq:S2rounding}
F^{(1)}_{\rm BWM} = e^{-\mathcal{S}_2} \sim e^{-s_{\rm eq} L_A} + e^{-v_2 s_{\rm eq} t} \;.
\end{equation}
This is easily generalised to $k>1$. Indeed, since the paths can be assumed to be non-interacting, the $k$ paths can independently choose whether to exit from the left or the bottom, i.e. $F^{(k)}_{\rm BWM}\sim [F^{(1)}_{\rm BWM}]^k$ for $k>1$. Consequently,
\begin{equation}
\label{eq:Deltak2}
    \left[ \Delta^{(k)} \right]^2 \sim 
    \left(1 + \exp[(-\tilde{v}_2 t +  L_A)s_{\rm eq} ]\right)^k - 1 
=:    \left[ \Delta^{(k)} \right]^2_{\rm non-int} 
    \;.
\end{equation}
Note that Eq.~\eqref{eq:Deltak2} describes the regime $t = O(L_A)$ where the contributions of the configurations with paths exiting the left or the bottom are comparable. In the regimes where $t \ll  L_A$ and $t \gg  L_A$, Eq.~\eqref{eq:Deltak2} is consistent with Eq.~\eqref{eq:Fkmembrane}.
\begin{figure}[h]
        \includegraphics[width=0.49\textwidth]{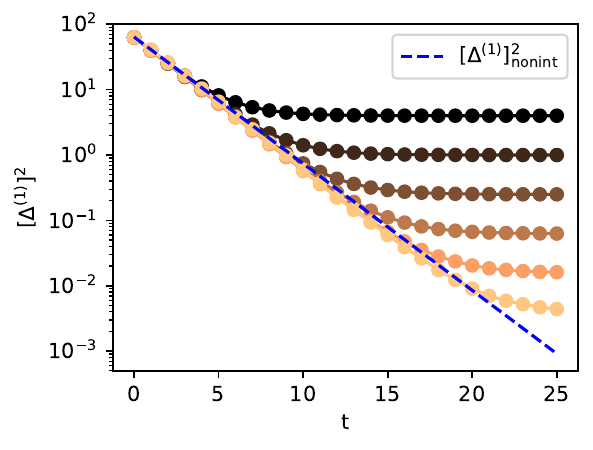}
    \hfill 
        \includegraphics[width=0.49\textwidth]{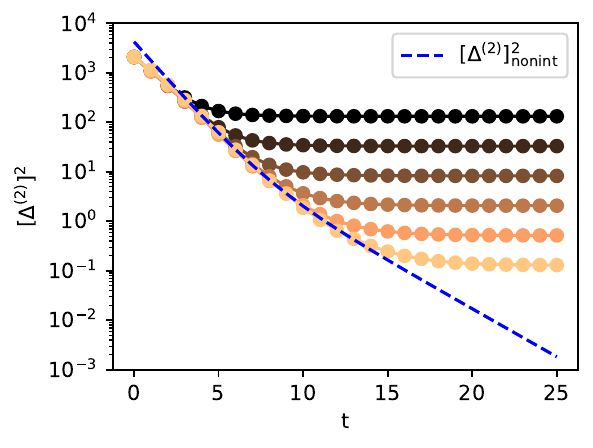}
    \hfill 
        \includegraphics[width=0.49\textwidth]{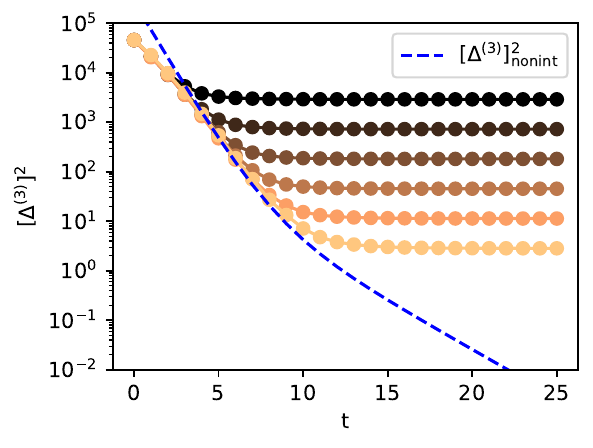}
    \caption{
    Comparison between numerical simulations for $q=2$ and $L_A = 6$ 
    of the distance $\Delta^{(k)}$ in Eq.~\eqref{eq:DeltakF} and the theoretical prediction in Eq.~\eqref{eq:Deltak2} with $k = 1,2,3$ and $L_B = 4,6,\ldots,16$ (black to orange). }
    \label{fig:comparison_Deltak2}
\end{figure}
\noindent We can now use Eq.~\eqref{eq:Deltak2} to estimate the scaling of design times. From Eq.~\eqref{eq:DeltakF}, we get
\begin{equation}
    \Delta^{(k)} < \epsilon \quad \Rightarrow (1 + \exp[(-v_2 t_k +  L_A)s_{\rm eq} ])^k  \lesssim 1 + \epsilon^2 \;.
\end{equation}
Expanding for $t_k \gtrsim L_A/v_2$, it leads to
\begin{equation}
    t_k \sim \frac{L_A}{v_2} + \frac{2 \ln(1/\epsilon)}{v_2 s_{\rm eq}}+ \frac{\ln(k)}{v_2 s_{\rm eq}} = t_1 + 
    \frac{\ln(k)}{v_2 s_{\rm eq}} \;,
\end{equation}
with a logarithmic dependence of the design times on $k$. This result is in agreement with what was found in \cite{ippoliti2022solvable} and shows that that the logarithmic behavior as a function of $k$ is rooted in the exponential suppression of the subdominant term in Eq.~\eqref{eq:S2rounding}, ultimately a consequence of the membrane picture.
A quantitative comparison of Eq.~\eqref{eq:Deltak2} with numerical calculations performed at $q = 2$ is shown in Fig.~\ref{fig:comparison_Deltak2}.

\section{Non-locality and order of limits}\label{sec:non_local}

\begin{figure}[h]
\begin{center}
\includegraphics[width=0.7\textwidth]{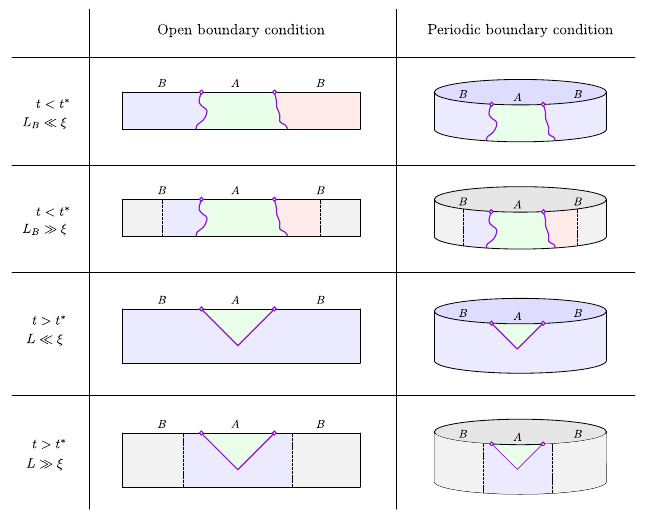}
\caption{
A table representing the dominant contribution to the frame potential for different boundary conditions.
%
For $t < t^*= L_A/v$ and $L_B \ll \xi$, there is a difference between the obc and pbc cases, since the left and right end of the configurations of obc can take different permutations, while the ones of pbc can only take 1 permutation in the region $B$.
For $t < t^*$ and $L_B \gg \xi$, the difference between the obc and pbc cases is suppressed as a function of $L_B$.
%
For $t>t^*$, the left and right regions of $B$ enclose $A$ with a domain wall. These $B$ regions become connected and takes an identical permutation.
\label{fig:obc_pbc}}
\end{center}
\end{figure}

It is worth discussing the membrane picture for the frame potentials of the projected ensemble in a modified protocol, where the region $A$ sits in the bulk of the system. This set-up was recently considered in~\cite{shrotriya2023nonlocality} in the context of dual unitary circuits. 
As illustrated in Fig.~\ref{fig:obc_pbc}, in the presence of open boundary conditions (obc), the complement $B$ of the region $A$ is actually composed by two disconnected regions. On the contrary, for periodic boundary conditions (pbc), $B$ is made up of a single connected component.
Let us analyze the consequence of this difference in boundary conditions from the perspective of the membrane picture as it emerges at $q \to \infty$ first. As explained in Sec.~\ref{sec:setup}, when the boundary state $\kket{B}$ is expanded in terms of permutation states, one gets several contributions from each spatial site weighted according to Eq.~\eqref{eq:Bsigma}. Then, the $q\to\infty$ limit has two effects: first, because of the form of Eq.~\eqref{eq:Bsigma}, it selects only factorised permutations with the form $\sigma = (\sigma_{n+k}, \sigma_{n+k}')$ such that $\mathbf{1}_F(\sigma) = 1$; second, a spatial variation of the pair $(\sigma_{n+k}, \sigma_{n+k}')$ within the region $B$ would lead to an additional membrane with a cost suppressed in $q$. For obc, one is free to choose two different pairs $(\sigma_{n+k}, \sigma_{n+k}')$ and $(\tilde\sigma_{n+k}, \tilde\sigma_{n+k}')$ that are spread throughout each connected component of $B$. This extra possibility for obc is associated with an enhancement of the frame potential at short times $t \ll L_A / (2 v_2)$, due to the independent sums over the permutations in each connected component of $B$, that is,
\begin{equation}
\label{eq:qinfobcpbc}
    F_{\rm BWM}^{(k)} \stackrel{q\to \infty}{=} 
    \begin{cases}
    (k!)^2 e^{-2 v_2 k t}&\text{obc}\;,\\
    k! e^{-2 v_2 k t}&\text{pbc} \;.
    \end{cases}
\end{equation}
Note the factor $2$ in the exponential decrease compared to Eq.~\eqref{eq:Fkmembrane} (recall that $v^{(k)} \sim k v_2$) that comes from the existence of two boundaries for the region $A$ in the current geometry.  

\begin{figure}[h]
        \includegraphics[width=0.85\textwidth]{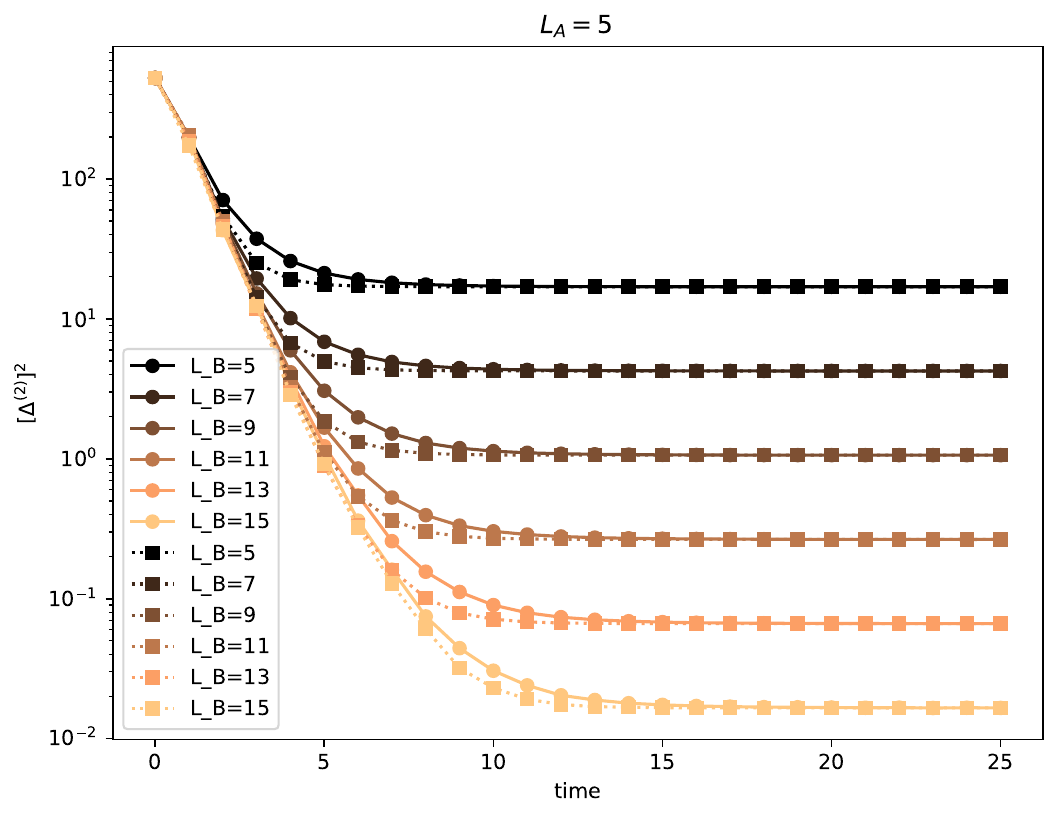}
    \caption{
    Numerical comparison between obc (circle solid line) and pbc (square dotted line) of the squared distance at $q=2$, $k=2$ for an interval of size $L_A = 5$ sitting in the middle of the chain and several values of $L_B$. No nonlocality is observed as for larger and larger $L_B$ the two boundary conditions fall one on top of the other.}
    \label{fig:comparison_pbc_obs}
\end{figure}

At large times, both boundary conditions lead to the same saturation value $F_{\rm BWM}^{(k)} \to F_{\rm Haar}^{(k)}$ as the region $B$ is eventually connected again throughout $A$ itself (see Fig.~\ref{fig:obc_pbc}, third and forth row). This behavior is particularly appealing, as it seems to survive even in the $L_B \to \infty$ limit, where one normally expects boundary conditions to play no role in standard thermodynamic quantities. In fact, an analogous type of enhancement of the frame potential at $k > 1$ with obc has been reported in \cite{shrotriya2023nonlocality} for dual unitary circuits and has been justified on the basis of the non-local effect of projective measurements in quantum mechanics~\cite{RevModPhys.81.1727}. Therefore, it is natural to ask whether, in the case of the BWM under consideration, an effect of the boundary conditions survives at finite $q$.
The question can be rephrased whether the region $B$ can support long-range order in the permutation basis. If $q$ is large but finite, we can consider the effect of adding an extra membrane (domain wall) in the bulk of the region $B$ that separates the pair $(\sigma_{n+k}, \sigma_{n+k}')$ from $(\tau \sigma_{n+k}, \sigma_{n+k}')$ or $(\sigma_{n+k}, \tau \sigma_{n+k}')$, where $\tau \in S_{n+k}$. For sufficiently large $L_B$, we can assume that it is entropically favorable to restrict to the elementary domain walls, that is, $\tau = \tau_{a,b}$, the transposition exchanging the elements $a$ and $b$ with $1\leq a<b\leq n$. Accounting for these two cases, there are $2 \times \binom{n + k}{2} = (n +k )(n+ k - 1)$ such  transpositions, while the cost associated to such a membrane is the same as the one determining the purity $\sim e^{-t v_2 s_{\rm eq}}$. Given that the membrane can be placed anywhere in the region $B$, we see that order will be lost when $L_B \gg \xi_{n,k}(t) \sim e^{v_2 s_{\rm eq} t} /[(n+k)(n+k-1)] $, where $\xi_{n,k}(t)$ represents the effective correlation length in the permutation space. If $q \to \infty$, $\xi_{n,k}(t)$ trivially diverges for all $n,k$ and $t$, confirming the scenario of Eq.~\eqref{eq:qinfobcpbc}. We expect that the fine-tuning required for dual-unitary circuits (considered in~\cite{shrotriya2023nonlocality}) formally leads to a divergent correlation length in our framework. 
On the contrary, at finite $q$, there is a nontrivial dependence on the ordering of limits between $L_B \to \infty$, required by the definition of the projected ensemble, and $n \to 1-k$ as prescribed by the replica trick \eqref{eq:replicalim} where $\xi_{n,k}(t) \to \infty$. However, we expect that this divergence in the correlation length is an unphysical consequence of having restricted our analysis only to diluted transpositions, an approximation that required $L_B$ to be large in the first place. So we expect that the consistent way to take these limits is to send $L_B \to \infty$ before the $n \to 1- k$ limit. In that case, the arguments here suggest that the long-range order is always destroyed and no significant difference appears between obc and pbc. The numerical simulation shown in Fig.~\ref{fig:comparison_pbc_obs} is consistent with this conclusion.

\section{Conclusions}\label{sec:con}
This work investigates the problem of deep thermalization in the context of random unitary circuits (RUC), by mapping the frame potential to a statistical mechanical problem.
We connect the calculation of the frame potential with the moments of state purity using the replica trick, revealing the emergence of a membrane as a key feature. 
Through our analysis, we estimate the time scales for the emergence of deep thermalization for the $k$-th moment  to be $t_k=O(\log k)$, a prediction that seems to be robust for generic chaotic quantum circuit as long as the membrane picture is applicable.
For outlooks, it would be of interest to understand the behaviour of deep thermalization in the presence of symmetries and conserved quantities~\cite{vonKeyserlingk2017a, Huse2017, friedman2019, mark2024maximumentropyprincipledeep}: In particular symmetry breaking can give rise to prethermalisation (or more generally slow thermalisation)~\cite{PhysRevLett.115.180601, Kormos2017} which is expected to be seen also in the higher-replica observables studied here. It is also of interest to understand the subleading-in-$q$ differences between deep thermalization and state purity by refining the replica analysis, which can perhaps be investigated using the non-Hermitian Ginibre ensemble via space-time duality \cite{Shivam_2023}.

\section{Acknowledgements}
We are thankful for Wen Wei Ho and Matteo Ippoliti for helpful discussions. AC acknowledges the support from the Royal Society grant RGS{$\backslash$}R1{$\backslash$}231444 and from the EPSRC Open Fellowship EP/X042812/1. ADL acknowledges support from the ANR JCJC grant ANR-21-CE47-0003 (TamEnt). 

\bibliography{biblio}

\newpage

\appendix
\section{Projected ensemble for a Haar-random state}\label{app:pe_haar}
Consider the situation where the system has been prepared in a Haar-distributed random state $\ket{\Psi}$. As explained in the main text, we use the replica trick to write the generalized frame potential in terms of $2(n + k)$ replicas and consider, in the end, the replica limit $n \to 1 - k$. Here, we work with arbitrary value of $q$. The Haar average over $\ket{\Psi}$ gives
\begin{equation}\label{app_eq:haar_state_id}
\overline{ \left( \ket{\Psi}\bra{\Psi} \right)^{\otimes (2n + 2k)}} = \frac{1}{\mathcal{N}_{2n+2k}(q^L)}\sum_{\sigma \in \mathcal{S}_{2n + 2k}} \kket{\sigma}^{\otimes L} \,,
\end{equation}
where we defined
\begin{equation}
\mathcal{N}_m(q) := q(q+1)(q+2)\ldots (q + m - 1) \,. 
\end{equation}
By using Eq.~\eqref{app_eq:haar_state_id} and a straightforward generalization of \eqref{eq:fkn}, and by applying the boundary conditions at the top in Fig.~\ref{fig:frame_rho}, we obtain
\begin{equation}
F^{(n,k)} = \frac{1}{\mathcal{N}_{2n+2k}(q^L)}\sum_{\sigma \in \mathcal{S}_{2n + 2k}} \bbraket{A|\sigma}^{L_A} \bbraket{B|\sigma}^{L_B} \,,
\end{equation}
For the sites in $B$, we have
\begin{equation}
\bbraket{B | \sigma} = \sum_{a,a'} \mathbf{1}_F(\sigma)  + (1 - \mathbf{1}_F(\sigma)) \delta_{a,a'} = 
\mathbf{1}_F(\sigma)(q^2 - q) + q \,. 
\end{equation}
where $\mathbf{1}_F(\sigma) = 1$ if $\sigma$ is factorized as $(\sigma_1, \sigma_2)$, with $\sigma_{1}, \sigma_{2} \in \mathcal{S}_{n+k}$ and $0$ otherwise. We also have
\begin{equation}
\bbraket{B|\sigma}^{L_B} = (q^{2L_B} - q^{L_B})\mathbf{1}_F(\sigma) + q^{L_B} \,,
\end{equation}
The sum over $\sigma$ involves two different terms. The first takes the form
\begin{equation}
\sum_{\sigma \in \mathcal{S}_{2n +2k}} \bbraket{A | \sigma}^{L_A} = \sum_{\sigma \in \mathcal{S}_{2n +2k}} q^{L_A N_c(\mu_A \sigma^{-1})} = \mathcal{N}_{2n + 2k}(q^{L_A}) \,,
\end{equation}
where, as in the main text, $N_c(\sigma)$ counts the number of cycles in the permutation $\sigma \in S_m$. The second contribution is
\begin{equation}
\sum_{\sigma \in \mathcal{S}_{2n +2k}} \bbraket{A | \sigma}^{L_A} \mathbf{1}_F(\sigma) = \sum_{\sigma_1,\sigma_2 \in \mathcal{S}_{n+k}}\bbraket{A | (\sigma_1, \sigma_2) }^{L_A} \,.
\end{equation}
To evaluate this last term, we notice that the permutation in $A$ acts as the identity over the first/last $n$ copies and it acts as $\imath_{2k}$ on the inner $2k$ copies (see Fig.~\ref{fig:frame_rho}). 
We can use this fact to effectively reduce the sums in $\mathcal{S}_{n+k}$ to sums in $\mathcal{S}_k$. To do so, we denote as $\left(\bbra{1_n} \otimes \mathbb{1}_k\right)$ the projector on the identity permutation for the first $n$ replicas, where $\mathbb{1}_k$ denotes the identity operator on the remaining $k$ replicas. Then, we have the identity 
\begin{equation}\label{app_eq:Z}
\sum_{\sigma \in \mathcal{S}_{n+k}}
\left[ \left(\bbra{1_n} \otimes \mathbb{1}_k\right)   \kket{\sigma} \right]^{L_A} = Z \sum_{\tilde{\sigma} \in \mathcal{S}_k} \kket{\tilde \sigma}^{\otimes L_A} \,.
\end{equation}
This equality can be simply understood noting that the left-hand side is invariant under any permatution of the last $k$ replicas and so it must be proportional to the flat sum over all elements in the group appearing in the right-hand side. The multiplicative constant $Z$ can be fixed by contracting both side over the identity on the residual $k$ copies,
\begin{equation}
\mathcal{N}_{n + k}(q^{L_A}) = \sum_{\sigma \in \mathcal{S}_{n+k}} q^{L_A N_c(\sigma)} = Z \sum_{\tilde{\sigma} \in \mathcal{S}_k} q^{L_A N_c(\tilde{\sigma}) } = Z \mathcal{N}_{k}(q^{L_A}) \quad \Rightarrow \quad Z = \frac{\mathcal{N}_{n + k}(q^{L_A})}{\mathcal{N}_{k}(q^{L_A})} \,.
\end{equation}
Substituting this back into \eqref{app_eq:Z}, we have
\be
\begin{aligned}
\sum_{\sigma \in \mathcal{S}_{2n +2k}} \bbraket{A | \sigma}^{L_A} \mathbf{1}_F(\sigma)  =& 
\left(\frac{\mathcal{N}_{n + k}(q^{L_A})}{\mathcal{N}_{k}(q^{L_A})}\right)^2 \sum_{\tilde{\sigma}_1, \tilde{\sigma}_2 \in \mathcal{S}_k} \bbraket{\imath_{2k} | (\tilde{\sigma}_1, \tilde{\sigma}_2) }^{L_A}
\\
=&
\left(\frac{\mathcal{N}_{n + k}(q^{L_A})}{\mathcal{N}_{k}(q^{L_A})}\right)^2 \sum_{\tilde{\sigma}_1, \tilde{\sigma}_2 \in \mathcal{S}_k} q^{L_A N_c( \tilde{\sigma}_1 \tilde{\sigma}_2^{-1})} 
=
 k! \frac{\mathcal{N}_{n + k}(q^{L_A})^2}{\mathcal{N}_{k}(q^{L_A})}  \,.
\end{aligned}
\ee
Putting everything together, we arrive at
\begin{equation}
F^{(n,k)} = \frac{1}{\mathcal{N}_{2n+2k}(q^L)} \left[q^{L_B} \mathcal{N}_{2n + 2k}(q^{L_A}) + (q^{2L_B} - q^{L_B})k! \frac{\mathcal{N}_{n + k}(q^{L_A})^2}{\mathcal{N}_{k}(q^{L_A})} \right] \,.
\end{equation}
Taking the limit $n \to 1 - k$, we have
\begin{multline}
F^{(k)} = \lim_{n \to 1 - k} F^{(n,k)} = \frac{1}{\mathcal{N}_{2}(q^L)} \left[(q^{L_B} \mathcal{N}_{2}(q^{L_A}) + (q^{2L_B} - q^{L_B})k! \frac{\mathcal{N}_{1}(q^{L_A})^2}{\mathcal{N}_{k}(q^{L_A})} \right] = \\=
\frac{1}{q^L(q^L + 1)} \left[(q^{L_B} q^{L_A}(q^{L_A}+1)+ \frac{(q^{2L_B} - q^{L_B})k! q^{2L_A}}{\mathcal{N}_{k}(q^{L_A})} \right]  \,.
\end{multline}
With $L = L_A + L_B$, in the limit $L_B \to \infty$, we have
\begin{equation}
F^{(k)} = \frac{k!}{\mathcal{N}_k(q^{L_A})} \,.
\end{equation}
\noindent Further, we can verify that the limits involved in $L_B \to \infty, n \to 1 - k$ and our approximation $q \to \infty$ commutes 
among themselves,
\begin{equation}
F^{(n,k)} \stackrel{L_B \to \infty}{=} k! \frac{q^{2L_B}}{q^{L(2n + 2k)}}\frac{\mathcal{N}_{n + k}(q^{L_A})^2}{\mathcal{N}_{k}(q^{L_A})} \stackrel{q \to \infty}{=} \frac{1}{q^{L_B(2n + 2k - 2)}}\frac{k!}{q^{k L_A}} \stackrel{n \to 1 -k}{=} \frac{k!}{q^{k L_A}}\,.
\end{equation}
Note that the large-$q$ expansion is formally equivalent to the large--$L_A$ expansion for a Haar-random state. However, for generic local quantum circuits, the time-dependence of the frame potential explicitly depends on $L_A$ even though $q \to \infty$.

\end{document}